\documentclass{article}
\pdfoutput=1

\usepackage{amssymb,amsfonts,amsmath}
\usepackage{cite,enumerate,float}
\usepackage{color}
\usepackage{tikz}
\usetikzlibrary{arrows,snakes,backgrounds}

\usepackage{amssymb,amsfonts,amsmath,stmaryrd}
\usepackage{cite,enumerate,float,indentfirst}
\usepackage{color}
\usepackage{tikz}
\usetikzlibrary{arrows,snakes,backgrounds,calc}
\usepackage{ytableau}
\usepackage[vcentermath]{youngtab}

\def\be{\begin{eqnarray}}
\def\ee{\end{eqnarray}}
\def\nn{\nonumber}

\def\be{\begin{eqnarray}}
\def\ee{\end{eqnarray}}
\def\nn{\nonumber}

\def\p{\partial}
\def\tr{{\rm tr}\,}
\def\Tr{{\rm Tr}\,}

\def\B{W}
\def\c{\pi}

\definecolor{red}{rgb}{1,0,0}
\definecolor{orange}{rgb}{1,0.5,0}
\definecolor{violet}{rgb}{0.7,0,1}

\hoffset= - 1.0in         

\begin{document}

\title{\vspace{1.5cm}\bf
Deformation of  superintegrability\\
in the Miwa-deformed Gaussian matrix model
}

\author{
A. Mironov$^{b,c,d,}$\footnote{mironov@lpi.ru,mironov@itep.ru},
A. Morozov$^{a,c,d,}$\footnote{morozov@itep.ru},
A. Popolitov$^{a,c,d,}$\footnote{popolit@gmail.com},
Sh. Shakirov$^{d,}$\footnote{shakirov.work@gmail.com}
}

\date{ }

\maketitle

\vspace{-6.5cm}

\begin{center}
\hfill FIAN/TD-03/24\\
\hfill IITP/TH-04/24\\
\hfill ITEP/TH-05/24\\
\hfill MIPT/TH-04/24
\end{center}

\vspace{4.5cm}

\begin{center}
$^a$ {\small {\it MIPT, Dolgoprudny, 141701, Russia}}\\
$^b$ {\small {\it Lebedev Physics Institute, Moscow 119991, Russia}}\\
$^c$ {\small {\it NRC ``Kurchatov Institute", 123182, Moscow, Russia}}\\
$^d$ {\small {\it Institute for Information Transmission Problems, Moscow 127994, Russia}}
\end{center}

\vspace{.1cm}

\begin{abstract}
We consider an arbitrary deformation of the Gaussian matrix model parameterized by Miwa variables $z_a$.
One can look at it as a mixture of the Gaussian and logarithmic (Selberg) potentials,
which are both superintegrable.
The mixture is not, still one can find an explicit expression for an arbitrary Schur average
as a linear transform of a {\it finite degree} polynomial made from the values of skew Schur functions
at the Gaussian locus $p_k=\delta_{k,2}$.
This linear operation includes multiplication with an exponential  $ e^{z_a^2/2}$
and a kind of Borel transform of the resulting product, which we call multiple and enhanced.
The existence of such remarkable formulas appears intimately related to the theory
of auxiliary $K$-polynomials, which appeared in {\it bilinear} superintegrable  correlators
at the Gaussian point (strict superintegrability). We also consider in great detail the generating function of
correlators $\langle(\Tr X)^k\rangle$ in this model, and discuss its integrable determinant representation. At last, we describe deformation of all results to the Gaussian $\beta$-ensemble.
\end{abstract}

\bigskip

\section{Introduction}

This paper is a direct extension of our recent discovery \cite{MMPS1}
that perturbation theory around a superintegrable point can be extraordinary simple
and can actually be summed up into an absolutely explicit formula
with the help of what we called an {\it enhanced} Borel transform.
In this paper, we develop this result in two directions.

First, we express perturbative corrections through the values of Schur functions
at the special point $p_k=\delta_{k,2}$, which makes similarity to {\it super}integrability
nearly full, and allows us to call the entire phenomenon {\it a deformation of superintegrability}.
The only difference is that now the {\it skew} Schur functions are involved.
In fact, these are rather sophisticated special functions, they are still polynomials,
but with non-trivial coefficients.
The formulas from \cite{MMPS1} are actually computationally better, but conceptually
an expression via the Schur functions has many advantages.

Second, we make use of one of advantages and describe various generalizations
like introducing many Miwa variables (and {\it multiple} Borel transform) and $\beta$-deformation.
A very interesting generalization to expansions around non-Gaussian monomial potentials
is now also straightforward \cite{MMMPS3}, but it involves a number of new ideas:
in addition to just shifting the locus to $p_k =\delta_{k,r}$, there is also an ambiguity
in the choice of superintegrability preserving integration contours among all the
(non-superintegrable) Dijkgraaf-Vafa phases.
We leave this story to a separate paper in order to avoid overloading
and shadowing the ideas of the present one.

The rest of the Introduction briefly repeats the main points from \cite{MMPS1},
since we need them below as well.

Superintegrability in quantum field theory was originally defined in \cite{IMM,MM} (based on the
phenomenon earlier observed in \cite{DiF}--\cite{Pop}, see also some preliminary results in \cite{Kaz}--\cite{MKR} and later progress in \cite{MMten}--\cite{MO},\cite{MMssi}--\cite{CMPT})
as a possibility of finding a basis in the space of correlators, when they can be
explicitly calculated.
The basis can be a somewhat transcendental (not expressed in elementary functions),
but just a little, unlike the case of generic non-superintegrable models.
This was supposed to mimic the situation in superintegrable classical potentials,
like the harmonic oscillator or Newton/Coulomb potentials, when the orbits become
periodic and are expressed in terms of periodic, though still elliptic integrals
(i.e. not just elementary trigonometric functions).
As a basic example in QFT, we took the Gaussian matrix model \cite{MM},
other examples include Selberg (logarithmic) models \cite{Sel}--\cite{Kad2},\cite{MMSh,MMShS} and a variety
of other theories \cite{MMsi}.
This basic example states that the Gaussian average of the Schur function\footnote{The Schur function $S_R(z_a)$ is a symmetric function of a set of variables $\{z_a\}$ or of their power sums $p_k=\sum_az_a^k$. In the latter case, we use the notation $S_R\{p_k\}$. The Schur function is labelled by the Young diagram (partition) $R$: $R_1\ge R_2\ge\ldots\ge R_{l_R}>0$, and can be manifestly evaluated using the Jacobi-Trudi formula
$$
S_R\{p_k\}=\det_{i,j}h_{R_i-i+j}\{p_k\}
$$
where the complete homogeneous symmetric polynomials $h_n\{p_k\}$'s are defined as
$$
\exp\Big(\sum_k{\frac{p_kz^k}{k}}\Big)=\sum_m{h_n\{p_k\}z^n}\nn
$$
}
$S_R$ is explicitly calculable
in terms of the same Schur functions:
\be\label{siG}
\Big< S_R \Big> = \eta_R(N) S_R\{\delta_{k,2}\}
\ee
with $\eta_R(N):= \frac{S_R\{N\}}{S_R\{\delta_{k,1}\}}$,
where the Schur function is labeled by the Young diagram $R$ and is a graded polynomial of time variables,
which is expressed at the l.h.s. through ``the quantum fields"  $p_k = \tr X^k$, $X$ being a
$N\times N$ Hermitian matrix in the case of Gaussian matrix model. The correlators in this latter are defined
\be\label{corG}
\Big<\ldots \Big> =  \int  dX e^{-\frac{1}{2}\tr X^2} \ \ldots
\ee
and are normalized in such a way that $\langle1\rangle=1$. The measure $dX$ is a natural flat Lebesgue measure (invariant under
translations) on Hermitian matrices. At the r.h.s. of the Schur function in (\ref{siG}), the variables are restricted to particular loci $p_k = \delta_{k,m}$ and $p_k = N$.
The ratio $\eta_R(N)$ is just a polynomial in $N$, a product over the boxes $(i,j)$ of the Young diagram:
$\eta_R = \prod_{(i,j)\in R} (N+i-j)$.

In this paper, we are going to consider a deformation of the Gaussian model\footnote{In the deformed case, we do not change the normalization of correlators as compared with the non-deformed case in order to have formulas simpler.}:
\be\label{dGm}
\Big<\ldots \Big>_{\c_k}=\int dX\exp\left(-{1\over 2}\Tr X^2+\sum_k{\c_k\over k}\Tr X^k\right) \ \ldots
\ee
The r.h.s. of this expression is understood as a power series in $\c_k$'s, i.e. this is an arbitrary deformation that is associated with the same integration contour along the real axis as in the Gaussian model. In other words, this is a small deformation over the Gaussian background, and parameters $\{\c_k\}$ are considered as describing small deviations from the Gaussian action. Here we consider only the specialization $\c_k=\pm z^k$ and explain the structure of correlators in such a model. It appears to have an interesting and intriguing feature: it turns out that the correlators are given by a Borel transform of a finite degree polynomial times a quadratic exponential.

In \cite[sec.3]{MMssi}, we used the model (\ref{dGm}) as a typical {\it counter}example for {\it superintegrability}.
Now we try to move further and explain how this notion can be extended and generalized
to capture this more sophisticated model. One of the most natural options is to choose the parameters $\c_k$'s in the Miwa parametrization: $\c_k = \sum_{i=1}^m z_i^k$. Not surprisingly, the quantity $m$ plays a distinguished role in the resulting formulas.
In this paper, we start from the simplest case of $m=1$ and explain the structure of correlators in such a model. It appears
to have an interesting and intriguing feature: it turns out that the correlators are given by a Borel transform of a finite degree polynomial times a quadratic exponential.

Let us explain the origin of this Borel transform. To this end, we notice that, in the Gaussian model, the generating function of all correlators looks like
\be
Z\{p_k\} :=  \sum_R S_R\{p_k\} \Big< S_R \Big>
= \sum_R  \eta_R(N) S_R\{\delta_{k,2}\}S_R\{p_k\}
\ee
If not this $\eta_R(N)$, one would get an elementary answer, using the Cauchy formula
\cite{Mac}:
\be
\sum_R    S_R\{\bar p_k \}S_R\{p_k\} = \exp\left(\sum_k \frac{\bar p_k p_k}{k}\right) \ \
\Longrightarrow \ \
\sum_R    S_R\{\delta_{k,2}\}S_R\{p_k\} = e^{p_2/2}
\ee
As explained in \cite{China1,China2}, insertion of the polynomial factor $\eta_R(N)$ can be induced
by an action of a linear operator acting on the time variables $p_k$:
\be
Z\{p_k\} = \widehat{\cal O}(N)\left\{ \sum_R    S_R\{\delta_{k,2}\}S_R\{p_k\}\right\} =
\widehat{\cal O}(N)  e^{p_2/2}
\ee
The action of operator $\widehat{\cal O}$ can be interpreted as an {\it enhanced} Borel transform.
Usually Borel ``improves" infinite series by inserting extra factorials \cite{Borel}
or their combinatorial counterparts \cite{MMBorel} in denominators.
In the case of polynomial insertions like $\eta_R$,
the series are sometimes\footnote{In the case of choosing the deformation of the Gaussian model by $\c_k=-\sum_az_a^k$, see sec.6 below.} {\it cut off} by a combination of $\Gamma$-function factors $\frac{N!}{(N-n)!}$
at finite values of $n$, actually regulated by the size $N$ of the matrix in
the underlying matrix model.
This cutoff is a kind of extreme (enhanced) version of the same convergency-improvement idea.

In order to demonstrate that emerging the Borel transform is a general feature of the Miwa-deformed model, in the paper, we consider a deformation with an arbitrary $m$. However, the correlators in this case become rather involved. This is why we specifically consider in detail arbitrary correlators of $\Big(\Tr X\Big)^k$ in this model with $m$ Miwa variables, and demonstrate that, in order to imitate $\eta_R(N)$, the kind of borelization is applied independently to all the Miwa variables, thus what we come across is actually
a {\it multiple} enhanced Borel transform.

Summing up series in $\c_k$ provides an example of {\it resurgence} \cite{Mar},
remarkable because it should one day give rise
to new superintegrable theories, e.g. for monomial potentials $\tr X^p$ with integer $p\geq 3$.
Like every non-trivial resurgence, it involves extra parameters like $p-2$ different phases
(Stokes sectors) for monomial potentials or Dijkgraaf-Vafa phases for non-monomial potentials.
This interesting direction is beyond the scope of the present paper, because Miwa deformations with
any finite number $m$ of $z_a$ variables do not bring us to a new superintegrable point with multiple
Stokes sectors.
Instead, we can treat these deformations as a kind of superintegrable {\it themselves},
because we found a way to express the answers for averages $\Big<S_R\Big>_{\pi_k=\sum z_a^k}$
with arbitrary $R$ through finite and explicit polynomials made from Schur functions. In fact, even more is correct:
as we demonstrated in \cite{MMPS1}, at least, in the case of one Miwa variable, all correlators are explicitly
expressed in terms of peculiar polynomials $\B_j({\bf Q})$
depending on somewhat mysterious truncation $Q\longrightarrow {\bf Q}$ of the Young diagram $Q$.
At least, this looks like quite a non-trivial and interesting generalization of superintegrability.

\bigskip

The paper is organized as follows. In section 2, we discuss a structure of correlators in the Gaussian model with general deformation, and in detail in the case of deformation by a single Miwa variables. In this latter case, we realize natural to introduce an enhanced Borel transform, which we define and discuss in section 3.  In section 4, we explain that instead of considering averages of the Schur functions, one can deal with another set of polynomials, which have simpler averages. This allows us to obtain a general formula for correlators in the Gaussian model deformed with arbitrary many Miwa variables and to extend it to the $\beta$-ensemble in section 5. In section 6, we consider correlators of $\Big(\Tr X\Big)^k$ in the Gaussian model deformed by $m$ Miwa variables, discuss its integrable properties in section 7, and its $\beta$-deformation in section 8. Section 9 contains some concluding remarks.

\paragraph{Notation.} We use the notation $S_R(z_a)$ and $S_R(z_1,z_2,\ldots,z_n)$ for the Schur functions, which are symmetric functions of a set of $n$ variables $\{z_a\}$, and $S_R\{p_k\}$ for the Schur functions as functions of power sums $p_k=\sum_az_a^k$, which are graded polynomials of variables $p_k$ of grading $k$. The Schur function $S_R\{p\}$ is labelled by the Young diagram (partition) $R$: $R_1\ge R_2\ge\ldots\ge R_{l_R}>0$, and the grading of this Schur function is $|R|:=\sum_iR_i$. We also denote through $S_{R/Q}\{p\}$ the skew Schur functions:
\be\label{skew}
S_R\{p_k+p'_k\}=\sum_QS_{R/Q}\{p_k\}S_Q\{p'_k\}
\ee
Similarly, we use notations $J_R(.)$ and $J_R\{.\}$ for the Jack polynomials \cite{Mac}.

Throughout the paper, we use the Pochhammer symbol
\be\label{Poch}
(N;\mu)_n=\prod_{k=0}^{n-1} (N+k\mu)
\ee
and $(N;\mu)_0=1$.

\section{Miwa deformation of superintegrability for Gaussian model\label{secMiwa1}}

Thus, in this paper, we consider the correlators in the deformed Gaussian model. The generating function of the correlators is given by
\be
Z=\int dX\exp\left(-{1\over 2}\Tr X^2+\sum_k{p_k+\c_k\over k}\Tr X^k\right)
\label{p+c}
\ee
and, using (\ref{siG}), we immediately obtain (see (\ref{skew}))
\be\label{Z}
Z=\sum_R\eta_R(N)S_R\{\delta_{k,2}\}S_R\{p_k+\c_k\}=\sum_{R,Q}\eta_R(N)S_R\{\delta_{k,2}\}S_{R/Q}\{\c_k\}S_Q\{p_k\}
\ee
i.e., for an arbitrary correlator, one obtains
\be\label{gc}
\left<S_Q\{\Tr X^k\}\right>_{\c_k}=\sum_R\eta_R(N)S_R\{\delta_{k,2}\}S_{R/Q}\{\c_k\}
\ee
This is an infinite sum. It is always divisible by to $\eta_Q$, since the summand is non-zero only if the Young diagram $R$ contains all boxes of the Young diagram $Q$ inside. Hence, the correlator can be written as
\be\label{11}
\left<S_Q\{\Tr X^k\}\right>_{\c_k}=\eta_Q(N)F_Q(\c_k,N)
\ee
where
\be\label{FQ}
F_Q(\c_k,N):=\sum_R\frac{\eta_R(N)}{\eta_Q(N)}S_R\{\delta_{k,2}\}S_{R/Q}\{\c_k\}
\label{FQdef}
\ee
is a power series in $N$ and $\c_k$. Here the (infinite) sum runs over all Young diagrams $R$ containing the diagram $Q$.

For $\c_k=0$ and $F_Q=S_R\{\delta_{k,2}\}$, one obtains the usual superintegrability formula (\ref{siG}) \cite{MMsi},
for $\c_k\neq 0$, one gets a {\it deformation}, which we can treat as that of superintegrability
provided $F_Q(\c_k,N)$ are calculable and simple enough.

As we already explained, the Miwa parametrization of the deforming constants $\c_k=\sum_iz_i^k$ looks quite natural. In this case, one
inserts in the Gaussian integral an additional factor of\footnote{One can avoid a potential singularity of the integral shifting $z_i$ from the real axis. In practice, we always deal with this factor as a power series in $z_i$.}
\be
\prod_{i=1}^m{1\over \det(1-z_iX)}
\ee
In the case of generic $m$, formula (\ref{FQ}) for $F_Q(\c_k,N)$ does not look simple (though it possesses an interesting structure that we discuss throughout the paper), however, in the case of $m=1$, it does:
\be\label{gc1}
F_Q(z,N)=\sum_Rz^{|R|-|Q|}\frac{\eta_R(N)}{\eta_Q(N)}S_R\{\delta_{k,2}\}S_{R/Q}\{1\}
\ee
where $|R|=\sum_iR_i$ denotes the size of the Young diagram $R$.

There is a systematic description of the correlators in the case of the Gaussian model deformed by one Miwa variable.
It turns out that one can evaluate the infinite sums, (\ref{FQ}) for $F_Q(z,N)$ in this case.

The simplest correlators in this case are
\be
F_{[2r]} = \sum_{k=0}^\infty \sum_{j=0}^{{\rm min}(k,r)} \frac{(N+2j;1)_{2k-2j}z^{2k}}{(2r-2j)!!(2k-2j)!!}
= \sum_{k=0}^\infty  \sum_{j=0}^{{\rm min}(k,r)} \frac{(N+2k-1)!}{(N+2j-1)!}\frac{z^{2k}}{(2r-2j)!!(2k-2j)!! } \nn \\
F_{[2r+1]} = \sum_{k=0}^\infty \sum_{j=0}^{{\rm min}(k,r)} \frac{(N+2j+1;1)_{2k-2j}z^{2k+1}}{(2r-2j)!!(2k-2j)!!}
= \sum_{k=0}^\infty  \sum_{j=0}^{{\rm min}(k,r)} \frac{(N+2k)!}{(N+2j)!} \frac{z^{2k+1}}{(2r-2j)!!(2k-2j)!!}
\ee
and
\be\label{13}
F_{[1^{2s}]} = \frac{(-)^s}{(2s)!!} \sum_{k=0}^\infty  \frac{(N+2k-1)!}{(N-1)!}\frac{z^{2k}}{(2k)!!} \nn \\
F_{[1^{2s+1}]} = \frac{(-)^s}{(2s)!!} \sum_{k=0}^\infty  \frac{(N+2k)!}{N!}\frac{z^{2k+1}}{(2k)!!}
\ee
The general answer for $F_Q$ of even size $|Q|$ is
\be\label{FQeven}
\boxed{
F_Q = \sum_{k=0}^\infty \frac{z^{2k}}{(2k)!!} \sum_{j=0}^k c_{j}(Q) \frac{k!}{(k-j)!}\frac{(N+2k-1)!}{(N+2j-1)!}=
\sum_{k=0}^\infty (N;1)_{2k}\frac{z^{2k}}{(2k)!!} \sum_{j=0}^k c_{j}(Q) \frac{(k;-1)_j}{(N;1)_{2j}}
}
\ee
while that for $F_Q$ of odd size $|Q|$ is
\be
\boxed{
F_Q =\sum_{k=0}^\infty \frac{z^{2k+1}}{(2k)!!} \sum_{j=0}^k c_{j}(Q) \frac{k!}{(k-j)!}\frac{(N+2k)!}{(N+2j)!}=
\sum_{k=0}^\infty (N;1)_{2k+1}\frac{z^{2k+1}}{(2k)!!} \sum_{j=0}^k c_{j}(Q) \frac{(k;-1)_j}{(N;1)_{2j+1}}
}
\label{FQodd}
\ee

Now we explain various ingredients of these formulas.

\paragraph{Overall coefficients.} An important observation is that if the constant term in $F_Q$ at even $|Q|$ or the linear term in $z$ in $F(Q)$ at odd $|Q|$, the whole $F_Q$ is zero, i.e. the corresponding coefficients are, in fact, the overall coefficients. Hence, we need to describe them in order to determine when $F_Q$ vanishes.

The overall coefficient in the case of even $|Q|$ in (\ref{FQeven}) follows immediately from (\ref{FQ}): the coefficient $c_0(Q)$ in front of $z^0$ is determined by $R=Q$ and is equal just to $S_Q\{\delta_{k,2}\}$. The value of $S_Q\{\delta_{k,2}\}$ is equal \cite{Pop,MMNO} to
\be\label{S2}
S_Q\{\delta_{k,2}\}=\delta_2(Q)\prod_{(i,j)\in Q}{1\over h_{i,j}^{ev}}
\ee
where $h_{i,j}$ is the hook length, and the product runs over only hook with even length, which we denoted by the superscript $ev$. $\delta_2(Q)$ is defined in \cite[Eq.(3.26)]{Pop}:
\be
\delta_2(Q)=\left\{
\begin{array}{cl}
(-1)^{|Q|/2}\prod_{(i,j)\in Q}(-1)^{[c_{i,j}/2]+[h_{i,j}/2]}&\hbox{ if the $2$-core of $Q$ is trivial}\cr\cr
0&\hbox{otherwise}
\end{array}
\right.
\ee
where $c_{i,j}$ is the content of the box $(i,j)$ in $Q$. Thus, $c_j(Q)=0$ if the $2$-core of $Q$ is non-trivial\footnote{In accordance with \cite{Mac}, the $r$-core of $Q$ is the unique result of stripping all possible rim hooks of length $r$
from $Q$.}.

The overall coefficients in the case of odd $|Q|$ is proportional to $z$, i.e. only the Young diagrams $R$ with $|R|=|Q|+1$ contributes to the sum (\ref{FQ}):
\be
c_0(Q)=\sum_{Q+\Box}(N+j_\Box-i_\Box)S_{Q+\Box}\{\delta_{k,2}\}
\ee
where the sum goes over all Young diagrams produced from $Q$ by adding a box, and $i_\Box$, $j_\Box$ denote the coordinates of the added box on this new Young diagram $Q+\Box$.
Note that this coefficient does not depend on $N$. Indeed, since \cite{Mac}
\be
p_1S_Q\{p_k\}=\sum_{Q+\Box}S_{Q+\Box}\{p_k\}
\ee
we immediately obtain
\be
\sum_{Q+\Box}S_{Q+\Box}\{\delta_{k,2}\}=0
\ee
Hence,
\be
c_0(Q)=\sum_{Q+\Box}(j_\Box-i_\Box)S_{Q+\Box}\{\delta_{k,2}\}
\ee
Note that this sum can be reproduced by the action of the differential operator
$\sum_kkp_{k+1}{\p\over\p p_k}$, \cite{China1}:
\be
c_0(Q)=\sum_{Q+\Box}(j_\Box-i_\Box)S_{Q+\Box}\{\delta_{k,2}\}=\sum_kkp_{k+1}{\p\over\p p_k}S_Q\{p_k\}\Big|_{p_k=\delta_{k,2}}={\p\over\p p_1}S_Q\{p_k\}\Big|_{p_k=\delta_{k,2}}
:= S'_Q\{\delta_{k,2}\}
\ee

\bigskip

\paragraph{Coefficients $c_j(Q)$.} The coefficients $c_j(Q)$ are expressed through auxiliary functions, which we manifestly described in the Appendix of \cite{MMPS1}. The manifest formulas for them are
\be
\boxed{
\begin{array}{c}
c_j(Q)=2^jS_{Q/[2j]}\{\delta_{k,2}\}\ \ \ \ \ \ \hbox{for even }|Q|\cr
\cr
c_j(Q)=2^jS_{Q/[2j+1]}\{\delta_{k,2}\}\ \ \ \ \ \ \hbox{for odd }|Q|
\end{array}
}
\label{cvsS}
\ee
These quantities are the mysterious quantities from \cite{MMPS1} that were denoted there $C_Q\cdot \B_j({\bf Q})$, and they were parameterized an auxiliary Young diagram ${\bf Q}$ (see \cite[formulas (38) and (51)]{MMPS1}).
The $\B_j({\bf Q})$-representation of \cite{MMPS1} for correlators has its own advantages, for example, it is easily computable (when ${\bf Q}$ is known),
in variance with the Schur polynomials, which get increasingly complicated for large partitions.
However, (\ref{cvsS}) are {\it conceptually} important and emphasize the fact that
{\bf superintegrability is deformed only slightly} once expressed in an appropriate form and variables.
We prove these formulas in sec.4, they appear related to the theory of exact {\it bilinear}
Gaussian correlators (strict superintegrability) \cite{MMssi,MMskew}, which originally could seem a little artificial,
but now acquires a spectacular {\it raison d'\^etre}.

An important additional property that allows one to identify (\ref{cvsS}) with formulas from \cite{MMPS1}  is that
\be
 S'_Q\{\delta_{k,2}\}= S_{Q/[1]}\{\delta_{k,2}\}
\ee

From here, it is just one step  towards the final formulas (\ref{B2}) and (\ref{B3}) for the averages (\ref{FQeven}) and (\ref{FQodd}).
It, however, involves one more ingredient, which we will name the {\it enhanced} Borel transform.

\section{Borel transform}

{\it A priori}  the Schur averages $F_Q(z,N)$ in the Gaussian model deformed with  one Miwa variable $z$
are represented by series (\ref{FQeven}) and (\ref{FQodd}) in powers of $z$ (sums over $k$),
with the coefficients having three properties:
\begin{itemize}
\item[(a)] they are themselves some series (sums over $j$),
\item[(b)] which are actually cut to polynomials by peculiar factorials $(N-j)!$ in denominators
in the case of choosing $\c_k=-\sum_az_a^k$, see sec.6 below,
\item[(c)] a truly non-trivial part of the coefficients is defined by a certain truncation of $Q$
\cite{MMPS1}.
\end{itemize}

In this paper, we found remarkable formulas (\ref{cvsS}) which resolve point (c):
now we know what are these coefficients.

The next step is to find an appropriate interpretation of (b),
and below in this section we suggest that the inverse factorials $(N-j)!$ are an enhanced
version of the ordinary inverse factorials $j!$, which defined the Borel transform.
While the Borel transform damped the large $j$ asymptotics of the series,
our coefficients can just {\it cut} them down to polynomials,
hence we call this {\it enhanced} Borel transform.

A pleasant bonus of this definition is that if one treats the $j$-sums as the enhanced Borel transform,
the entire sum over $k$ can be explicitly summed up into the exponential $e^{z^2/2}$
times a polynomial!
Thus, by finding an interpretation for (b), we are simultaneously handling (a):
the final answer for the average $F_Q$ is the enhanced Borel transform of a simple
exponential times a polynomial.

In result, the averages are explicitly calculable,i.e., in a sense, are superintegrable,
just instead of a single Schur product, the answer is the enhanced Borel transform of a finite sum.

Moreover, items (a) and (b) can be immediately continued to several Miwa variables,
only one needs a {\it multiple} enhanced Borel transform.
One can sometimes avoid multiplicity in the transform, at expense of making it more sophisticated
like in \cite{MMMPS3} but, at the moment, the multiple option looks more preferable.

In the next sections, we also consider in detail the evaluation of
an arbitrary correlator $\Big<(\Tr X)^k\Big>$ in the Gaussian potential
perturbed by an arbitrary number of Miwa variables, $\c_k=\pm\sum_az_a^k$,
and explain what the {\it multiple enhanced Borel transform} is,
which converts it in a convenient form.
The generalization of point (c) is also straightforward, see the next section 4.

\subsection{Borel representation of correlators}

Let us introduce a Borel transformation defined in the following way: given a power series
\be\label{is}
F(z)=\sum_kF_kz^k
\ee
the Borel transformation (parameterized by a parameter $N$) is the power series
\be\label{oBor}
{\cal B}_N\Big[F(z)\Big]_z=\sum_k(N;1)_kF_kz^k
\ee
This transformation at negative integer $N$ regularizes the infinite series (\ref{is}) making a finite polynomial from it, hence the name (as we explained in the Introduction).

One may notice that all the correlators in the single Miwa deformed model possess such a form: (\ref{FQeven}), (\ref{FQodd}). Hence, all of them can be written the Borel transformation of the finite degree polynomial multiplied with the quadratic exponential. For instance, the results (\ref{13}) of the previous section can be rewritten using the Borel transform:
\be\label{B1}
F_{[1]}(z,N)=\sum_{k=0}(N+1;1)_{2k}{z^{2k+1}\over (2k)!!}={1\over N}{\cal B}_{N}^{(1)}\left[ze^{z^2\over 2}\right]_z\nn\\
\ee
while (\ref{FQeven}) has the form
\be\label{B2}
F_Q =
\sum_{k=0}^\infty (N;1)_{2k}\frac{z^{2k}}{(2k)!!} \sum_{j=0}^k c_{j}(Q) \frac{(k;-1)_j}{(N;1)_{2j}}
=\nn\\
={\cal B}_N\left[\sum_{k=0}^\infty \sum_{j=0}c_{j}(Q){(k;-1)_j\over (N;1)_{2j}}\frac{z^{2k}}{(2k)!!}\right]_z=
\boxed{
{\cal B}_N\left[e^{z^2\over 2}\sum_j{c_{j}(Q)\over (N;1)_{2j}}\left({z^2\over 2}\right)^j\right]_z}
\ee
and (\ref{FQodd}) has the form
\be\label{B3}
F_Q =
\sum_{k=0}^\infty (N;1)_{2k+1}\frac{z^{2k+1}}{(2k)!!} \sum_{j=0}^k c_{j}(Q) \frac{(k;-1)_j}{(N;1)_{2j+1}}=\boxed{
{\cal B}_N\left[ze^{z^2\over 2}\sum_j{c_{j}(Q)\over (N;1)_{2j+1}}\left({z^2\over 2}\right)^j\right]_z}
\ee
Here we have the Borel transform of finite degree polynomials times $e^{z^2\over 2}$.

The notable difference between (\ref{B1}) and (\ref{B2})-(\ref{B3}) is that, in the first of them, the answer is given by the Borel transform of a function that does not depend on $N$.

An essential point is that {\it all} correlators in the single Miwa deformed model can be presented as the Borel transform. This is a peculiarity of the model: in the multiple Miwa deformed model,  in order to describe all the correlators one has to generalize the Borel transform to the multiple Borel transform, which we discuss in this section.
In the next sections, we consider in detail evaluating an arbitrary correlator $(\Tr X)^k$ in the example of multiple Miwa deformation, $c_k=\pm\sum_az_a^k$ and demonstrate that they can be presented in a convenient form with the multiple Borel transform.

\subsection{The operator $\hat {\cal O}(N)$}

Let us construct the operator $\hat {\cal O}(N)$, which was called rotation operator in \cite{China2} and which plays the central role in the definition of the Borel transform. This operator is constructed manifestly via generalized cut-and-join operators. That is,
the generalized cut-and-join operators $\hat W_\Delta$ form a commutative set of operators, the Schur functions being their eigenfunctions \cite{MMN1}:
\be\label{caj}
\hat W_\Delta\ S_R=\phi_R(\Delta)\ S_R
\ee
where, for the diagram $\Delta$ containing $r$ unit cycles: $\Delta=[\tilde\Delta, 1^r]$,
\be\label{cphi}
\phi_{R}(\Delta)=\left\{\begin{array}{cl}
0\ \ \ \ \ \ \ \ \ \ \ \ \ \ \ \ \ \ \ \ \ \ \ \ \ \  &|\Delta|>|R|\\
&\\
\displaystyle{(|R|-|\Delta|+r)!\over r!(|R|-|\Delta|)!}\ \phi_{R}(\hat\Delta)=\displaystyle{(|R|-|\Delta|+r)!\over r!(|R|-|\Delta|)!}\
\displaystyle{\psi_{R}(\hat\Delta)\over z_{\hat\Delta}S_R\{\delta_{k,1}\}}\ \ \ \ \ \ \ \ \ \ \ \ \ \ \ \ \ \ \ \ \ \ \ &|\Delta|\le |R|
\end{array}\right.
\ee
Here $\hat\Delta:=[\Delta,1^{|R|-|\Delta|}]$, $\psi_R(\Delta)$ is the value of character of the permutation group ${\cal S}_{|R|}$ in the representation $R$ on the conjugacy class $\Delta$,  and $z_\Delta$ is the order of automorphism of the Young diagram $\Delta$ \cite{Fulton}. Now note that \cite[Eq.(61)]{MMN}
\be
\sum_\Delta \phi_R(\Delta)p_\Delta
= {S_R\{p_k+\delta_{k,1}\}\over d_R}
\ee
Now we construct the rotation operator as follows:
\be
\hat {\cal O}(u):=\sum_\Delta p_\Delta\cdot\hat W_\Delta,\ \ \ \hbox{with} \ p_k=u-\delta_{k,1}
\ee
Here we use the notation $p_\Delta=\prod_{i=1}^{l_\Delta}p_{\delta_i}$, where $l_\Delta$ is the length of the partition $\Delta$, and $\delta_i$'s are its parts.

Then,
\begin{equation}
    \hat {\cal O}(u) \cdot S_R\{p_k\} = \dfrac{S_R\{u\}}{S_R\{\delta_{k,1}\}} S_R\{p_k\}=
    \eta_R(u)\  S_R\{p_k\}
\end{equation}
This operator was constructed earlier in \cite[Eq.(21)]{AMMN} in order to insert additional factors $\dfrac{S_R\{N\}}{S_R\{\delta_{k,1}\}}$ into character expansion of the partition function, and was written there in a different form. In particular, one can rewrite it via Casimir operators \cite[sec.3]{AMMN}.

\subsection{Borel transform}

Now we define the Borel transform of an arbitrary symmetric power series $F(z_i,z_2,\ldots,z_m)$ of $m$ variables, which we sometimes call the $m$-fold (or multiple) Borel transform:
\be
\boxed{
{\cal B}_N^{(m)}\Big[F(z_i,z_2,\ldots,z_m)\Big]_z=\hat {\cal O}(N)\cdot F(z_i,z_2,\ldots,z_m)
}
\ee
In fact, the superscript $m$ is not that necessary, and we often omit it.

Let us explain this definition more explicitly. An arbitrary symmetric power series can be expanded into the basis of the Schur functions $S_R$, and, for $m$ variables, the Young diagrams $R$ can not have more than $m$ lines. Thus, one can generally write
\be
F(z_1,z_2,\ldots,z_m)=\sum_{n_1\ge n_2\ge\ldots\ge n_m\ge 0}
C_{n_1,n_2,\ldots,n_m}S_{[n_1,n_2,\ldots,n_m]}(z_1,z_2,\ldots,z_m)
\ee
and the $m$-fold Borel transform gives rise to
\be
{\cal B}^{(m)}_N\Big[F(z_1,z_2,\ldots,z_m)\Big]_z=\nn
\ee
\be
=\sum_{n_1\ge n_2\ge\ldots\ge n_m\ge 0}
(N;1)_{n_1}(N-1;1)_{n_2}\ldots(N-m+1;1)_{n_m}C_{n_1,n_2,\ldots,n_m}S_{[n_1,n_2,\ldots,n_m]}(z_1,z_2,\ldots,z_m)
\ee
Note that one can similarly use the relation
\be\label{Scon}
S_R\{p_k\}=S_{R^\vee}\{(-1)^{k+1}p_k\}
\ee
where $R^\vee$ denotes the transposed Young diagram, in order to get the expansion
\be
F(z_1,z_2,\ldots,z_m)=\sum_{n_1\ge n_2\ge\ldots\ge n_m\ge 0}
(-1)^{\sum n_i}C_{n_1,n_2,\ldots,n_m}S_{[n_1,n_2,\ldots,n_m]^\vee}\{-\sum_{i=1}^m z_i^k\}
\ee
and, in these terms,
\be
{\cal B}^{(m)}_{N}\Big[F(z_1,z_2,\ldots,z_m)\Big]_z=\nn
\ee
\be
=\sum_{n_1\ge n_2\ge\ldots\ge n_m\ge 0}(-1)^{\sum n_i}
(N;-1)_{n_1}(N+1;-1)_{n_2}\ldots(N+m-1;-1)_{n_m}C_{n_1,n_2,\ldots,n_m}
S_{[n_1,n_2,\ldots,n_m]^\vee}\{-\sum_{i=1}^m z_i^k\}=\\
=\sum_{n_1\ge n_2\ge\ldots\ge n_m\ge 0}
(-N;1)_{n_1}(-N-1;1)_{n_2}\ldots(-N-m+1;1)_{n_m}C_{n_1,n_2,\ldots,n_m}
S_{[n_1,n_2,\ldots,n_m]^\vee}\{-\sum_{i=1}^m z_i^k\}\nn
\ee
In particular, in the case of one variable, this transformation reduces to the ordinary enhanced Borel transform (\ref{oBor})
\be
\sum_k C_kz^k\stackrel{{\cal B}^{(1)}_{N}}{\longrightarrow}\sum_k (N;1)_kC_kz^k
\ee

\section{Averages in the Gaussian model with generic Miwa deformation
\label{fromK} }

We observed in the previous sections that the averages of the Schur functions in the Gaussian model deformed by a single Miwa variable are calculable and can be presented as the enhanced Borel transform of a polynomial times quadratic exponential. Let us explain that there is another basis \cite{MMssi,MMNek} instead of the Schur one, where these polynomials are simpler. Instead, the polynomials that form this basis are more complicated.

These polynomials are defined as
\be
K_R\{\Tr H^k\}=(-1)^{|R|}e^{\frac{1}{2}\tr H^2}\, S_R\{\hat W^-_k(H)\}\, e^{-\frac{1}{2}\tr H^2}
\ee
where\footnote{By the matrix derivative, we imply the derivative w.r.t. matrix elements of the transposed matrix: $\left(\frac{\partial}{\partial \Lambda}\right)_{ij}=\frac{\partial}{\partial \Lambda_{ji}}$.}
\be\label{W-}
\hat W^-_k(H)=\Tr\left( \frac{\p}{\p H}\right)^k
\ee
and $H$ is an $N\times N$ Hermitian matrix. The time variables are $p_k=\Tr H^k$. One can also realize the operators $\hat W^-_k$ as differential operators in $p_k$ from the very beginning \cite{MMMP1} so that
\be\label{Kp}
K_R\{p_k\}=(-1)^{|R|}e^{\frac{1}{2}p_2}\, S_R\{\hat W^-_k(p_k)\}\, e^{-\frac{1}{2}p_2}
\ee
However, this realization is quite involved.

The polynomials $K_R$ form a complete polynomial basis, however, they are not graded\footnote{Generalizations of these polynomials for other cases can be found in \cite[sec.7.3]{MMskew} and in \cite{CMPT}.}. The averages of these polynomials in the (non-deformed) Gaussian model celebrate two important (and defining) properties:
\be\label{KK}
\Big<K_R\cdot K_Q\Big>_G=\eta_R(N)\delta_{RQ}
\ee
and
\be\label{KS}
\Big<K_Q\cdot S_{R}\Big> _G=\eta_R(N)S_{R/Q}\{\delta_{k,2}\}
\ee
The second property is called strict superintegrability, since the bilinear correlator also turns out to be factorized. It follows from (\ref{KS}) that the average of $K_R$ in the deformed Gaussian model (\ref{dGm}) is
\be
\Big<K_Q\Big> _{\c_k}=\sum_RS_R\{\c_k\}\Big<K_Q\cdot S_R\Big> _G=\sum_R\eta_R(N)S_R\{\c_k\}S_{R/Q}\{\delta_{k,2}\}
\ee
This formula, in a sense, is dual to (\ref{gc}),
\be
\left<S_Q\right>_{\c_k}=\sum_R\eta_R(N)S_R\{\delta_{k,2}\}S_{R/Q}\{\c_k\}
\ee
but, in the case of $\c_k$ given by $m$ Miwa variables, can be immediately transformed to
\be
\boxed{
\Big<K_Q\Big> _{(m)}=\sum_R\eta_R(N)S_R\{z_a\}S_{R/Q}\{\delta_{k,2}\}={\cal B}_N\left[ \sum_RS_R\{z_a\}S_{R/Q}\{\delta_{k,2}\}\right]_z\stackrel{\hbox{\cite[p.71]{Mac}}}{=}{\cal B}_N\left[ S_Q\{z_a\}e^{{1\over 2}\sum_az_a^2}\right]_z
}
\ee
One can definitely expand the Schur functions over the polynomials $K_R$ and use the transition matrix in order to evaluate the average $\Big<S_Q\Big> _{(m)}$:
\be
S_Q=\sum_R C^R_QK_R\Rightarrow
\left\{
\begin{array}{c}
\Big<S_Q\Big> _{(m)}={\cal B}_N\left[\sum_R C^R_QS_R\{z_a\}e^{{1\over 2}\sum_az_a^2}\right]_z\cr
\sum_R \Big(C^{-1}\Big)^R_Q\Big<S_R\Big> _{(m)}={\cal B}_N\left[ S_Q\{z_a\}e^{{1\over 2}\sum_az_a^2}\right]_z
\end{array}
\right.
\ee
One can easily calculate the transition matrix from formulas (\ref{KK}) and (\ref{KS}), it turns out to be rather simple and triangle:
\be
C_Q^R={\eta_Q(N)\over\eta_R(N)}S_{Q/R}\{\delta_{k,2}\}
\ee
Its inverse is also simple
\be
\Big(C^{-1}\Big)_Q^R={\eta_Q(N)\over\eta_R(N)}S_{Q/R}\{-\delta_{k,2}\}
\ee
and, hence, one finally obtains
\be
\Big<S_Q\Big> _{(m)}={\cal B}_N\left[\sum_R {\eta_Q(N)\over\eta_R(N)}S_{Q/R}\{\delta_{k,2}\}S_R(z_a)e^{{1\over 2}\sum_az_a^2}\right]_z
\ee
and
\be
\boxed{
F_Q^{(m)}={\cal B}_N\left[\sum_R {S_{Q/R}\{\delta_{k,2}\}\over\eta_R(N)}S_R(z_a)e^{{1\over 2}\sum_az_a^2}\right]_z
}
\label{main}
\ee
which is a polynomial of degree (grading) $|Q|$ in $z_a$.

Using relation (\ref{Scon}), one can also obtain similar expressions for the choice of Miwa variables $\c_k=-\sum_{a=1}^mz_a^m$:
\be\label{mainn}
\boxed{
F_Q^{(m)}={\cal B}_{-N}\left[\sum_R (-1)^{|R|}{S_{Q/R^\vee}\{\delta_{k,2}\}\over\eta_{R^\vee}(N)}S_R(z_a)e^{-{1\over 2}\sum_az_a^2}\right]_z
}
\ee

In particular, at $m=1$, (\ref{main}) gives rise to expressions (\ref{FQodd}) and (\ref{FQeven}) coinciding with \cite[Eqs.(59)-(60)]{MMPS1} with the coefficients $c_j(Q)$ as in (\ref{cvsS}):
\be
c_j(Q)=2^jS_{Q/[2j]}\{\delta_{k,2}\}\ \ \ \ \ \ \hbox{for even }|Q|\nn\\
c_j(Q)=2^jS_{Q/[2j+1]}\{\delta_{k,2}\}\ \ \ \ \ \ \hbox{for odd }|Q|
\ee
Similarly, (\ref{mainn}) at $m=1$ gives rise to expressions coinciding with \cite[Eqs.(63)-(64)]{MMPS1} taking into account that
\be
S_{R^\vee}\{\delta_{k,2}\}\cdot S_{R/[1^{2n}]}\{\delta_{k,2}\}=(-1)^nS_{R}\{\delta_{k,2}\}S_{R^\vee/[2n]}\{\delta_{k,2}\}\nn\\
S'_{R^\vee}\{\delta_{k,2}\}\cdot S_{R/[1^{2n+1}]}\{\delta_{k,2}\}=(-1)^nS'_{R}\{\delta_{k,2}\}S_{R^\vee/[2n+1]}\{\delta_{k,2}\}
\ee
 We point out again that the normalization of the coefficients $c_j(Q)$ here and in \cite{MMPS1} differs by a factor of $S_R\{\delta_{k,2}\}$ in the case of even $|Q|$, and by a factor of $S'_R\{\delta_{k,2}\}$, in the case of odd $|Q|$.

Thus, we demonstrated that, in the case of the Gaussian model deformed with $m$ Miwa variables, the general correlators are reduced to infinite sums over $m$ integers  and only finite sums over Young diagrams instead of an infinite sum over Young diagrams as in (\ref{gc}). In the case of one Miwa variable, they are the infinite sums (\ref{FQeven}), (\ref{FQodd}) over integer $k$ and finite sums over integer $j$ instead of an infinite sum over the Young diagram $R$ in (\ref{gc1}).

\section{Deformed $\beta$-ensemble}

The construction of the previous section can be easily extended to the deformed Gaussian $\beta$-ensemble, where the correlators are defined
\be
\Big<\ldots \Big>^\beta_{\c_k} =  \int\prod_i dx_i \Delta^{2\beta}(x)e^{-\frac{1}{2}\sum_ix_i^2+\sum_{i,k}{\c_k x_i^k\over k}} \ \ldots
\ee
instead of (\ref{corG}), where $\Delta(x)=\prod_{i<j}(x_i-x_j)$ is the Vandermonde determinant.

In this case, the proper basis of polynomials is formed by the Jack polynomials \cite{Mac} instead of the Schur functions since they provide the superintegrability basis in this case \cite{MMS,Max}, and the operators $\hat W^-_k(H)$ (\ref{W-}) are now substituted by operators $\hat W^{(-,\beta)}_k$, which can be formulated either again as a differential operator in variables $p_k=\sum_ih_i^k$, or in terms of eigenvalues $h_i$ of the matrix $H$: in this case, they are just Hamiltonians of the rational Calogero model with the particle coordinates $h_i$ \cite{MMCal},\cite[sec.7]{MMMP1}. A possibility of this $\beta$-deformation is in no way a surprise since the construction is described by the so called integer ray Hamiltonians of \cite{MMMP1}, which are immediately $\beta$-deformed \cite{MMMP2}.

In order to describe the operators $\hat W^{(-,\beta)}_k$ in terms of variables $p_k$, we introduce a $\beta$-deformed counterpart of the operator $\hat {\cal O}(N)$ \cite{China2}, which is defined
\be
    \hat {\cal O}^\beta(N) \cdot J_R\{p_k\} =  \eta_R^\beta(N)\  J_R\{p_k\}\nn\\
    \eta_R^\beta(N)={J_R\{N\}\over J_R\{\delta_{k,1}\}}=\prod_{i,j\in N}(N+\beta^{-1}(j-1)-i+1)
\ee
Then, one has
\begin{equation}\label{Wk}
\hat W^{(-,\beta)}_k= \Big(\hat {\cal O}^\beta(N)\Big)^{-1}\left( {k\over\beta}\dfrac{\partial}{\partial p_k}\right) \hat {\cal O}^\beta(N)
\end{equation}
so that formula (\ref{Kp}) is now substituted with
\be
K_R^{(\beta)}\{p_k\}=(-1)^{|R|}e^{\frac{1}{2}p_2}\, J_R\left\{\hat W^{(-,\beta)}_k(p_k)\right\}\, e^{-\frac{1}{2}p_2}
\ee

Now, following \cite{China2}, we use that
\be\label{sJ}
J_Q\left\{{k\over\beta}{\p\over\p p_k}\right\}\ J_R\{p_k\}=||J_Q||\cdot J_{R/Q}\{p_k\}
\ee
where $||J_Q||$ is the norm square of the Jack polynomial,
\be
||J_R||:={\overline{G}^\beta_{R^\vee R}(0)\over G^\beta_{RR^\vee}(0)}\beta^{|R|}\ \ \ \ \ \ \
G_{R'R''}^\beta(x):=\prod_{(i,j)\in R'}\Big(x+R'_i-j+\beta(R''_j- i+1)\Big)
\ee
with the bar over the functions denoting the substitution $\beta\to\beta^{-1}$, use that
\be
J_{R/P}=\sum_Q\bar N^{R^\vee}_{Q^\vee P^\vee}\ J_Q
\ee
where where $N^Q_{RP}$ are the generalized Littlewood-Richardson coefficients for the Jack polynomials \cite{Mac},
and also use the superintegrability of the $\beta$-ensemble \cite{MMS,Max}
\be
\Big< J_R\Big>^\beta_G= \eta_R^\beta(N)J_R\{\beta\delta_{k,2}\}
\ee
in order to obtain
\begin{equation}
\begin{split}
         &\left< K^{(\beta)}_{P}\cdot J_Q\{p_k\} \right>^\beta_Q=
         \left< J_{P}\left\{\hat W^{(-,\beta)}_k\right\}J_Q\{p_k\} \right>^\beta_G  =
      \left< \Big(\hat {\cal O}^\beta(N)\Big)^{-1} \cdot J_P\left\{{k\over\beta}\dfrac{\partial}{\partial p_k} \right\} \cdot\hat {\cal O}^\beta(N) \cdot J_Q\{p_k\} \right>^\beta_G = \\
      &=\eta_Q^\beta(N) \left< \Big(\hat {\cal O}^\beta(N)\Big)^{-1}J_P\left\{{k\over\beta}\dfrac{\partial}{\partial p_k} \right\}
      J_Q\{p_k\} \right>^\beta_G
      =\eta_Q^\beta(N) ||J_P||\left< \Big(\hat {\cal O}^\beta(N)\Big)^{-1}J_{Q/P}\{p_k\} \right>^\beta_G = \\
      &=\eta_Q^\beta(N) ||J_P||\sum_Q\bar N^{Q^\vee}_{R^\vee P^\vee}\left< \Big(\hat {\cal O}^\beta(N)\Big)^{-1}J_{R}\{p_k\} \right>^\beta_G
      =  \eta_Q^\beta(N) ||J_P||\sum_Q\bar N^{Q^\vee}_{R^\vee P^\vee}{1\over \eta_R^\beta(N)}\Big< J_{R}\{p_k\} \Big>^\beta_G=\\
      &= \eta_Q^\beta(N) ||J_P||\sum_Q\bar N^{Q^\vee}_{R^\vee P^\vee} J_R \{\beta\delta_{k,2}\}=
        \eta_Q^\beta(N) ||J_P|| J_{Q/P} \{\beta\delta_{k,2}\}
\end{split}
\end{equation}
From this expression, one obtains instead of (\ref{KK}) and (\ref{KS}):
\be\label{KKb}
\Big<K^{(\beta)}_R\cdot K^{(\beta)}_Q\Big>^\beta_G=\eta_R^\beta(N) ||J_R||\delta_{RQ}
\ee
and
\be\label{KSb}
\Big<K^{(\beta)}_Q\cdot J_{R}\Big>^\beta _G= \eta_R^\beta(N) ||J_Q|| J_{R/Q} \{\beta\delta_{k,2}\}
\ee
Now, using the Cauchy identity
\be
\sum_R{J_R\{p_k\}J_R\{p'_k\}\over ||J_R||}=\exp\left(\beta\sum_k{p_kp'_k\over k}\right)
\ee
one can literally repeat all calculations of the previous section:
\be
\Big<K^{(\beta)}_Q\Big>^\beta _{\c_k}=\sum_R{J_R\{\c_k\}\over||J_R||}\Big<K^{(\beta)}_Q\cdot J_R\Big> _G=\sum_R\eta^\beta_R(N){||J_Q||\over||J_R||}J_R\{\c_k\}J_{R/Q}\{\beta\delta_{k,2}\}
\ee
This formula, in the case of $\c_k$ given by $m$ Miwa variables, can be immediately transformed to
\be
\boxed{
\Big<K^{(\beta)}_Q\Big>^\beta _{(m)}=\sum_R\eta^\beta_R(N){||J_Q||\over||J_R||}J_R(z_a)J_{R/Q}\{\beta\delta_{k,2}\}
={\cal B}^{(\beta)}_N\left[ \sum_R{||J_Q||\over||J_R||}J_R(z_a)J_{R/Q}\{\beta\delta_{k,2}\}\right]_z={\cal B}^{(\beta)}_N\left[ ||J_Q||J_Q(z_a)e^{{\beta\over 2}\sum_az_a^2}\right]_z
}
\ee
where we now define the enhanced multiple Borel transform with the action by the operator $\hat {\cal O}^\beta(N)$.

At last, one can calculate the transition matrix from formulas (\ref{KKb}) and (\ref{KSb}), it is again rather simple and triangle:
\be
J_R=\sum_Q\left({\eta^\beta_R(N)\over\eta^\beta_Q(N)}J_{R/Q}\{\beta\delta_{k,2}\}\right)\cdot K^{(\beta)}_Q
\ee
and, hence, one finally obtains
\be
\Big<J_Q\Big> ^\beta_{(m)}={\cal B}^{(\beta)}_N\left[\sum_R {\eta^\beta_Q(N)\over\eta^\beta_R(N)}J_{Q/R}\{\beta\delta_{k,2}\}||J_R||J_R(z_a)e^{{\beta\over 2}\sum_az_a^2}\right]_z
\ee
Defining similarly to (\ref{11})
\be
\Big< J_R\Big>^\beta_{(m)}= \eta_R^\beta(N)F_Q^{(m,\beta)}
\ee
one finds
\be
\boxed{
F_Q^{(m,\beta)}={\cal B}^{(\beta)}_N\left[\sum_R {J_{Q/R}\{\beta\delta_{k,2}\}\over\eta^\beta_R(N)}||J_R||J_R(z_a)
e^{{\beta\over 2}\sum_az_a^2}\right]_z
}
\ee
which is a polynomial of degree (grading) $|Q|$ in $z_a$.

\section{Gaussian model with linear term in Miwa variables}

Now we consider the Gaussian model deformed by the Miwa variables and the generating function of correlators of the form $\Big<(\Tr X)^k\Big>$, which is equivalent to adding the linear term to the Gaussian exponential. After integrating over angular variables, this partition function can be presented in the form
\be\label{ZN}
Z_N^{(\pm m)}={1\over \prod_{k=1}^Nk!}\int_{\infty}^{+\infty}{dw_1\over\sqrt{2\pi}}\ldots\int_{\infty}^{+\infty}{dw_N\over\sqrt{2\pi}}
\prod_{i<j}(w_i-w_j)^2e^{-{w_i^2\over 2}+tw_i+\sum_{i,k}{\c_kw_i^k\over k}}=\nn\\
\stackrel{\c_k=\pm\sum_az_a^k}{=}\ {1\over \prod_{k=1}^Nk!}\int_{\infty}^{+\infty}{dw_1\over\sqrt{2\pi}}\ldots\int_{\infty}^{+\infty}{dw_N\over\sqrt{2\pi}}
\prod_{i<j}(w_i-w_j)^2\prod_{a=1}^m\prod_{i=1}^N(1-z_aw_i)^{\mp 1}e^{-{w_i^2\over 2}+tw_i}
\ee
One can look differently at this partition function, and generate correlators in the Gaussian model with the linear term expanding this integral in $z_a$. We now demonstrate that the partition function $Z_N^{(\pm m)}$ is given by the Borel transform of a relatively simple function, if one parts off a peculiar  $N$-dependent factor.

\subsection{One Miwa variable $m=1$}

We start with the simplest example of one Miwa variable.

First of all, we expand
\be
\prod_{i=1}^N(1-zw_i)^{\pm 1}=\exp\left(\mp\sum_{i,k}{z^kw_i^k\over k}\right)=\sum_RS_R\{\mp z^k\}S_R\left\{\sum_iw_i^k\right\}
\ee
so that
\be
Z_N^{(\pm 1)}={1\over \prod_{k=1}^Nk!}\int_{\infty}^{+\infty}{dw_1\over\sqrt{2\pi}}\ldots\int_{\infty}^{+\infty}{dw_N\over\sqrt{2\pi}}
\prod_{i<j}(w_i-w_j)^2\left(\sum_RS_R\{\pm z^k\}S_R\left\{\sum_iw_i^k\right\}\right)e^{-{w_i^2\over 2}+tw_i}=\nn\\
=\sum_RS_R\{\pm z^k\}\left<S_R\left\{\sum_iw_i^k\right\}\right>
\ee
and use the superintegrability property of the Gaussian model with a linear shift \cite{Max,MMsi}\footnote{We point out that the partition function in these references is normalized with an additional factor of $e^{Nt^2/2}$.}:
\be\label{SI}
\left<S_R\left\{\sum_iw_i^k\right\}\right>=e^{Nt^2\over 2}\eta_R(N)S_R\left\{t\delta_{k,1}+\delta_{k,2}\right\}
\ee
Thus,
\be
Z_N^{(\pm 1)}=\sum_RS_R\{\pm z^k\}\left<S_R\left\{\sum_iw_i^k\right\}\right>=
e^{Nt^2\over 2}\sum_Rz^{|R|}\eta_R(N)S_R\{\pm 1\}S_R\left\{t\delta_{k,1}+\delta_{k,2}\right\}
\ee
Now note that $S_R\{1\}=\delta_{R,[n]}$ and, hence, $S_R\{-1\}=(-1)^n\delta_{R,[1^n]}$, and that
that
\be
S_{[k]}\left\{t\delta_{k,1}+\delta_{k,2}\right\}=\left[e^{z^2/2+zt}\right]_{k}
\ee
and, hence,
\be
(-1)^kS_{[1^k]}\left\{t\delta_{k,1}+\delta_{k,2}\right\}=\left[e^{-z^2/2-zt}\right]_{k}
\ee
Here $[\ldots ]_k$ denotes the coefficient in front of $z^k$, and this kind of formulas follows from the identity
\be
\sum_nS_{[n]}\{p_k\}z^n=\exp\left(\sum_k{p_kz^k\over k}\right)
\ee
Thus, finally, one obtains
\be
Z_N^{(\pm 1)}=e^{Nt^2\over 2}\sum_{k=0}(N;\pm 1)_{k}\left[e^{\pm z^2/2\pm zt}\right]_{k}z^k
=e^{Nt^2\over 2}\sum_{k=0}(\pm N;1)_{k}\left[e^{\pm z^2/2\pm zt}\right]_{k}(\pm z)^k
\ee
i.e.
\be\label{Z1}
\boxed{
Z_N^{(\pm 1)}=e^{Nt^2\over 2}{\cal B}_{\pm N}\left[e^{\pm z^2/2+ zt}\right]_{\pm z}
}
\ee

\subsection{$m$ Miwa variables\label{4.2}}

One can immediately generalize this calculations for arbitrary $\mu$. That is, one expands
\be
\prod_{a=1}^m\prod_{i=1}^N(1-z_aw_i)^{\pm 1}=\exp\left(\mp\sum_{i,k,a}{z_a^kw_i^k\over k}\right)=\sum_RS_R\left\{\mp \sum_az_a^k\right\}S_R\left\{\sum_iw_i^k\right\}
\ee
so that
\be
Z_N^{(\pm 1)}=\sum_RS_R\left\{\pm \sum_az_a^k\right\}\left<S_R\left\{\sum_iw_i^k\right\}\right>
\ee
and, using (\ref{SI}),
\be
Z_N^{(\pm m)}=\sum_RS_R\left\{\pm \sum_az_a^k\right\}\left<S_R\left\{\sum_iw_i^k\right\}\right>=
e^{Nt^2\over 2}\sum_R\eta_R(N)S_R\left\{\pm \sum_az_a^k\right\}S_R\left\{t\delta_{k,1}+\delta_{k,2}\right\}
\ee

\subsubsection{Evaluating $Z_N^{(m)}$}

Let us first evaluate the partition function with positive $m$:
\be\label{26}
Z_N^{(m)}=\sum_RS_R\left\{\sum_az_a^k\right\}\left<S_R\left\{\sum_iw_i^k\right\}\right>=
e^{Nt^2\over 2}\sum_R\eta_R(N)S_R\left(z_1,z_2,\ldots,z_m\right)S_R\left\{t\delta_{k,1}+\delta_{k,2}\right\}
\ee
Note that the Schur function $S_R$ that depends on $m$ variables is non-zero only if $R$ has no more than $m$ lines. Hence, we can replace the sum over all $R$ in this formula with the sum over $m$ lengths of lines $n_1\ge n_2\ge\ldots\ge n_m\ge 0$:
\be
Z_N^{(m)}=e^{Nt^2\over 2}
\sum_{n_1\ge n_2\ge\ldots\ge n_m\ge 0}
(N;1)_{n_1}(N-1;1)_{n_2}\ldots(N-m+1;1)_{n_m}\times\nn\\
\times S_{[n_1,n_2,\ldots,n_m]}(z_1,z_2,\ldots,z_m)
S_{[n_1,n_2,\ldots,n_m]}\left\{t\delta_{k,1}+\delta_{k,2}\right\}
\ee
In order to evaluate the last multiplier, we use the Macdonald scalar product \cite{Mac},\footnote{The contour integral here includes the factor ${1\over 2\pi i}$, and the integral surrounds zero so that this is just the residue at the origin.}
\be\label{Shsp}
{1\over m!}\oint\prod_{i=1}^m{d\xi_i\over \xi_i}S_R(\xi)S_Q(\xi^{-1})\prod_{i\ne j}\left(1-{\xi_i\over \xi_j}\right)=\delta_{RQ}
\ee
and the Cauchy identity
\be\label{Cauchy}
\sum_RS_R\{p_k\}S_R(\xi_1,\ldots)=e^{\sum_{k,a}{p_k\xi_a^k\over k}}
\ee
i.e.
\be
\sum_{n_1\ge n_2\ge\ldots\ge n_m\ge 0}S_{[n_1,n_2,\ldots,n_m]}\left\{t\delta_{k,1}+\delta_{k,2}\right\}
S_{[n_1,n_2,\ldots,n_m]}(\xi_1,\xi_2,\ldots,\xi_m)=e^{t\sum_a\xi_a+{1\over 2}\sum_a\xi_a^2}
\ee
Thus, we finally obtain
\be\label{Zm1}
Z_N^{(m)}={1\over m!}e^{Nt^2\over 2}
\sum_{n_1\ge n_2\ge\ldots\ge n_m\ge 0}
(N;1)_{n_1}(N-1;1)_{n_2}\ldots(N-m+1;1)_{n_m}S_{[n_1,n_2,\ldots,n_m]}(z_1,z_2,\ldots,z_m)\times\nn\\
\times\oint\prod_{a=1}^m{d\xi_a\over \xi_a}\prod_{a\ne b}\left(1-{\xi_a\over \xi_b}\right)
e^{t\sum_a\xi_a+{1\over 2}\sum_a\xi_a^2}S_{[n_1,n_2,\ldots,n_m]}(\xi_1^{-1},\xi_2^{-1},\ldots,\xi^{-1}_m)
\ee
Using the Cauchy identity (\ref{Cauchy}) with $p_k=\sum_az_a^k$, one can immediately rewrite this formula as the Borel transform:
\be\label{Zmf}
Z_N^{(m)}={1\over m!}e^{Nt^2\over 2}{\cal B}_N^{(m)}
\left[\oint\prod_{a=1}^md\xi_a{\prod_{a\ne b}(\xi_a-\xi_b)\over \prod_{a,b}(\xi_a-z_b)}e^{t\sum_a\xi_a+{1\over 2}\sum_a\xi_a^2}
\right]_z
\ee
where the integration contours surround the poles.

In the case of $m=1$, we immediately return to formula (\ref{Z1}).

\subsubsection{Evaluating $Z_N^{(-m)}$}

For evaluating $Z_N^{(-m)}$, we use formula (\ref{Scon}) in order to obtain instead of (\ref{26})
\be
Z_N^{(-m)}=\sum_RS_R\left\{-\sum_az_a^k\right\}\left<S_R\left\{\sum_iw_i^k\right\}\right>=
e^{Nt^2\over 2}\sum_R(-1)^{|R|}\eta_R(N)S_{R^\vee}\left(z_1,z_2,\ldots,z_m\right)S_R\left\{t\delta_{k,1}+\delta_{k,2}\right\}
\ee
so that
\be
Z_N^{(-m)}=e^{Nt^2\over 2}
\sum_{n_1\ge n_2\ge\ldots\ge n_m\ge 0}(-1)^{\sum_an_a}
(N;-1)_{n_1}(N+1;-1)_{n_2}\ldots(N+m-1;-1)_{n_m}\times\nn\\
\times S_{[n_1,n_2,\ldots,n_m]}(z_1,z_2,\ldots,z_m)
S_{[n_1,n_2,\ldots,n_m]^\vee}\left\{t\delta_{k,1}+\delta_{k,2}\right\}
\ee
Using the same way to evaluate the factor $S_{[n_1,n_2,\ldots,n_m]^\vee}\left\{t\delta_{k,1}+\delta_{k,2}\right\}
$ as in the previous subsection, but with the Cauchy identity in the form
\be
\sum_{n_1\ge n_2\ge\ldots\ge n_m\ge 0}S_{[n_1,n_2,\ldots,n_m]^\vee}\left\{t\delta_{k,1}+\delta_{k,2}\right\}
S_{[n_1,n_2,\ldots,n_m]}(\xi_1,\xi_2,\ldots,\xi_m)=\nn\\
\sum_{n_1\ge n_2\ge\ldots\ge n_m\ge 0}(-1)^{\sum_an_a}S_{[n_1,n_2,\ldots,n_m]}\left\{t\delta_{k,1}-\delta_{k,2}\right\}
S_{[n_1,n_2,\ldots,n_m]}(\xi_1,\xi_2,\ldots,\xi_m)=e^{t\sum_a\xi_a-{1\over 2}\sum_a\xi_a^2}
\ee
we obtain
\be\label{Zm2}
Z_N^{(-m)}={1\over m!}e^{Nt^2\over 2}
\sum_{n_1\ge n_2\ge\ldots\ge n_m\ge 0}
(N;-1)_{n_1}(N+1;-1)_{n_2}\ldots(N+m-1;-1)_{n_m}S_{[n_1,n_2,\ldots,n_m]}(z_1,z_2,\ldots,z_m)\times\nn\\
\times\oint\prod_{a=1}^m{d\xi_a\over \xi_a}\prod_{a\ne b}\left(1-{\xi_a\over \xi_b}\right)
e^{t\sum_a\xi_a-{1\over 2}\sum_a\xi_a^2}S_{[n_1,n_2,\ldots,n_m]}(\xi_1^{-1},\xi_2^{-1},\ldots,\xi^{-1}_m)
\ee
One again can rewrite this formula as the Borel transform:
\be\label{Zmf2}
Z_N^{(-m)}={1\over m!}e^{Nt^2\over 2}{\cal B}_{-N}^{(m)}
\left[\oint\prod_{a=1}^md\xi_a{\prod_{a\ne b}(\xi_a-\xi_b)\over \prod_{a,b}(\xi_a+z_b)}e^{t\sum_a\xi_a-{1\over 2}\sum_a\xi_a^2}
\right]_{-z}
\ee
and, again, in the case of $m=-1$, we immediately return to formula (\ref{Z1}).

\subsubsection{Moral}

Thus, we obtained the generating functions of correlators $\Big<(\Tr X)^k\Big>$ in the Gaussian model deformed by the Miwa variables as the Borel transform of a relatively simple function:
\be\label{Borelr}
\boxed{
Z_N^{(\pm m)}={1\over m!}e^{Nt^2\over 2}{\cal B}_{\pm N}^{(m)}
\left[\oint\prod_{a=1}^md\xi_a{\prod_{a\ne b}(\xi_a-\xi_b)\over \prod_{a,b}(\xi_a\mp z_b)}e^{t\sum_a\xi_a\pm {1\over 2}\sum_a\xi_a^2}\right]_{\pm z}
}
\ee
Technically, this formula is, in a sense, more convenient than the sums over all possible Young diagrams since they are just multiple ($m$-fold) sums of the Schur functions with no more than $m$ lines. These latter can be evaluated as $m\times m$ determinants:
\be
S_{[n_1,n_2,\ldots,n_m]}(z_1,z_2,\ldots,z_m)={\det_{m\ge i,j\ge 1}z_i^{m+n_j-j}\over\Delta(z)}
\ee
where $\Delta(z)=\prod_{i<j}(z_i-z_j)$ is the Vandermonde determinant.

Moreover, at integer $N$, the infinite series $Z_N^{(-m)}$ becomes a finite sum: this is the main role of the Borel transform.

\section{Integrability and determinant representation}

It is known that the partition function of the Hermitian matrix model (\ref{p+c}) at any constants $\c_k$'s is a $\tau$-function of the integrable hierarchy: the (forced) Toda chain with the size of matrix $N$ playing the role of the discrete (zeroth) time, and $t_k=p_k/k$ being the remaining time variables of the hierarchy \cite{GMMMO,KMMOZ}. The Toda chain $\tau$-function satisfies the bilinear differential Hirota equations \cite{JM,DJKM,UT}, and, after the Miwa transformation of variables\footnote{Some of $\zeta_i$ may coincide giving rise to non-unit multiplicities. However, it is a smooth procedure, and, for the sake of simplicity, we consider here only unit multiplicities.}, $p_k=-\sum_az_a^k=-\sum_a \zeta_a^{-k}$, it satisfies the bilinear difference Hirota equations \cite{Miwa,GKM,Kharchev}. Moreover, the $\tau$-function in these variables always has the determinant representation
\be
\tau_N={\det_{a,b}\psi_a(\zeta_b)\over\Delta(\zeta)}
\ee
where $\psi_i(\zeta)=\zeta^{i-1}\Big(1+O(\zeta^{-1})\Big)$ form the Segal-Wilson basis describing some point in the
Sato’s Grassmannian \cite{GKM}. The concrete form of $\psi_i(\zeta)$ for the Hermitian matrix model at any constants $\c_k$'s was obtained in \cite{versus}:
\be
\psi_a(\zeta)=\zeta^{-N}P_{N+a-1}(\zeta)
\ee
where $P_a(\zeta)=\zeta^a+\ldots$ are polynomials in $\zeta$ orthogonal with the weight $e^{-{1\over 2}\zeta^2 +\sum_k{\c_k\zeta^k\over k}}$. In the case of model (\ref{ZN}), when $\c_k=t\delta_{k,1}$, these polynomials are the Hermite polynomials
\be
P_a(\zeta)=He_a(\zeta-t)=a!\sum_{j=0}{(-1)^j\over j!(a-2j)!}{(\zeta-t)^{a-2j}\over 2^j}=
a!\sum_{j=0}\sum_{k=0}{(-1)^j\over  (a-2j-k)!j!k!}{\zeta^{a-2j-k}t^k\over 2^j}
\ee
where $[\ldots]$ denotes the integer part of a number. Thus,
\be
\psi_a(\zeta)=
\sum_{j=0}\sum_{k=0}(-1)^j{(N+a-1)!\over  (N+a-1-2j-k)!j!k!}{\zeta^{a-1-2j-k}t^k\over 2^j}=
\zeta^{a-1}{\cal B}_{-N-a+1}\left[e^{-t\zeta^{-1}-{\zeta^{-2}\over 2}}\right]_{\zeta^{-1}}
\ee
and this finally gives us the determinant representation for the partition function (\ref{ZN}) (choosing the proper normalization),
\be
\boxed{
Z_N^{(-m)}={e^{Nt^2\over 2}\over m!}\cdot{\det_{a,b}\left(z_b^{1-a}{\cal B}_{-N-a+1}\left[e^{-tz_b-{z_b^2\over 2}}\right]_{z_b}\right)\over\Delta(z^{-1})}
}
\ee
which is equivalent to (\ref{Zmf2}).

\section{$\beta$-deformation}

The Gaussian matrix model after integrating over angular variables (\ref{ZN}) admits a simple deformation to the $\beta$-ensemble:
\be\label{be}
Z_{N,\beta}^{(\pm m)}=\left(\prod_{m=1}^{N}{\Gamma(\beta+1)\over (2\pi)^{1/2} \Gamma(m\beta+1)}\right)\ \int_{\infty}^{+\infty}{dw_1\over\sqrt{2\pi}}\ldots\int_{\infty}^{+\infty}{dw_N\over\sqrt{2\pi}}
\prod_{i<j}(w_i-w_j)^{2\beta}e^{-{\beta w_i^2\over 2}+\beta tw_i+\sum_{i,k}{p_kw_i^k\over k}}
\ee
Note that we preserve here $\beta$ positive in order to have the integral (\ref{be}) convergent.

In this case, the calculations of section 4 are literally repeated, with the following changes:
\begin{itemize}
\item The Miwa variables are now introduced with non-unit multiplicity: $\c_k=\beta\sum_az_a^k$. Hence, the integral (\ref{ZN}) becomes
\be
Z_{N,\beta}^{(\pm m)}=\left(\prod_{m=1}^{N}{\Gamma(\beta+1)\over (2\pi)^{1/2} \Gamma(m\beta+1)}\right)\ \int_{\infty}^{+\infty}{dw_1\over\sqrt{2\pi}}\ldots\int_{\infty}^{+\infty}{dw_N\over\sqrt{2\pi}}
\prod_{i<j}(w_i-w_j)^{2\beta}\prod_{a=1}^m\prod_{i=1}^N(1-z_aw_i)^{\mp \beta}e^{-{\beta w_i^2\over 2}+\beta tw_i}
\ee
\item The natural expansion is now into the Jack polynomials \cite{Mac}:
\be
\prod_{i=1}^N(1-zw_i)^{\pm \beta}=\exp\left(\mp\sum_{i,k}{\beta z^kw_i^k\over k}\right)=\sum_R{J_R\{\mp z^k\}J_R\left\{\sum_iw_i^k\right\}\over ||J_R||}
\ee
where we used the Cauchy identity for the Jack polynomials,
\be
\sum_R{J_R\{p_k\}J_R(\xi_1,\ldots)\over ||J_R||}=e^{\sum_{k,a}{\beta p_k\xi_a^k\over k}}
\ee
where
\be
||J_R||:={\overline{G}^\beta_{R^\vee R}(0)\over G^\beta_{RR^\vee}(0)}\beta^{|R|}\ \ \ \ \ \ \
G_{R'R''}^\beta(x):=\prod_{(i,j)\in R'}\Big(x+R'_i-j+\beta(R''_j- i+1)\Big)
\ee
\item The superintegrability property now reads
\be
\left<J_R\left\{\sum_iw_i^k\right\}\right>=e^{\beta Nt^2\over 2}\eta_R^{(\beta)}(N)J_R\left\{t\delta_{k,1}+\delta_{k,2}\right\}
\ee
where
\be
\eta_R^{(\beta)}(N)={J_R\{N\}\over J_R\{\delta_{k,1}\}}=\prod_{i,j\in R} (N+(j-1)\beta^{-1}-i+1)
\ee
\item The transposition rule (\ref{Scon}) for the Jack polynomials deforms to
\be
J_R\{\beta^{-1}p_k\}=||J_R||\cdot\overline{J}_{R^\vee}\{(-1)^{k+1} p_k\}
\ee
where the bar over any function denotes the replace $\beta\to\beta^{-1}$. In particular,
\be
||J_R||=||\overline{J}_{R^\vee}||^{-1}
\ee
\item The Macdonald scalar product (\ref{Shsp}) is deformed \cite{Mac} to
\be
{1\over m!}\oint\prod_{i=1}^m{d\xi_i\over \xi_i}J_R(\xi_i)J_Q(\xi^{-1}_i)\prod_{i\ne j}
\left(1-{\xi_i\over \xi_j}\right)^\beta={\cal N}_R\delta_{RQ}
\ee
with
\be
{\cal N}_R=\prod_{1\le i<j\le m}{\Gamma(R_i-\beta i-R_j+\beta j +\beta)\Gamma(R_i-\beta i-R_j+\beta j -\beta+1)\over
\Gamma(R_i-\beta i-R_j+\beta j )\Gamma(R_i-\beta i-R_j+\beta j +1)}
\ee
\item At last, the operator $\hat {\cal O}(N)$ also admits a natural $\beta$-deformation $\hat {\cal O}^{(\beta)}(N)$ \cite{China2},
\be
\hat {\cal O}^{(\beta)}(N)\cdot J_R=\eta_R^{(\beta)}(N)\cdot J_R
\ee
so that one now naturally defines the $\beta$-deformed multiple Borel transform
\be
{\cal B}_{N,\beta}^{(m)}\Big[F(z_i,z_2,\ldots,z_m)\Big]_z=\hat {\cal O}^{(\beta)}(N)\cdot F(z_i,z_2,\ldots,z_m)
\ee
which acts on the symmetric function expanded into the basis of the Jack functions,
\be
F(z_1,z_2,\ldots,z_m)=\sum_{n_1\ge n_2\ge\ldots\ge n_m\ge 0}
C_{n_1,n_2,\ldots,n_m}J_{[n_1,n_2,\ldots,n_m]}(z_1,z_2,\ldots,z_m)
\ee
as
\be
{\cal B}^{(m)}_{N,\beta}\Big[F(z_1,z_2,\ldots,z_m)\Big]_z=\nn
\ee
\be
=\sum_{n_1\ge n_2\ge\ldots\ge n_m\ge 0}
(N;\beta^{-1})_{n_1}(N-1;\beta^{-1})_{n_2}\ldots(N-m+1;\beta^{-1})_{n_m}C_{n_1,n_2,\ldots,n_m}S_{[n_1,n_2,\ldots,n_m]}(z_1,z_2,\ldots,z_m)
\ee
\end{itemize}

\bigskip

\paragraph{Answers for the $\beta$-ensemble partition functions.}
Now one can repeat calculations of section 4, using these modifications, in order to obtain
\be
Z_{N,\beta}^{(m)}={1\over m!}e^{\beta Nt^2\over 2}
\sum_{n_1\ge n_2\ge\ldots\ge n_m\ge 0}
{1\over {\cal N}_{[n_1,n_2,\ldots,n_m]}}(N;\beta^{-1})_{n_1}(N-1;\beta^{-1})_{n_2}\ldots(N-m+1;\beta^{-1})_{n_m}
\times\nn\\
\times J_{[n_1,n_2,\ldots,n_m]}(z_1,z_2,\ldots,z_m)
\oint\prod_{a=1}^m{d\xi_a\over \xi_a}\prod_{a\ne b}\left(1-{\xi_a\over \xi_b}\right)^\beta
e^{\beta t\sum_a\xi_a+{\beta\over 2}\sum_a\xi_a^2}J_{[n_1,n_2,\ldots,n_m]}(\xi_1^{-1},\xi_2^{-1},\ldots,\xi^{-1}_m)
\ee
and
\be\label{Zm3}
Z_{N,\beta}^{(-m)}={1\over m!}e^{\beta Nt^2\over 2}
\sum_{n_1\ge n_2\ge\ldots\ge n_m\ge 0}{||\overline{J}_{{[n_1,n_2,\ldots,n_m]}^\vee}||\over \overline{\cal N}_{[n_1,n_2,\ldots,n_m]}}
(N;-1)_{n_1}(N+\beta^{-1};-1)_{n_2}\ldots(N+\beta^{-1}(m-1);-1)_{n_m}\times\nn\\
\times J_{[n_1,n_2,\ldots,n_m]}(z_1,z_2,\ldots,z_m)
\oint\prod_{a=1}^m{d\xi_a\over \xi_a}\prod_{a\ne b}\left(1-{\xi_a\over \xi_b}\right)^{\beta^{-1}}
e^{t\sum_a\xi_a-{1\over 2}\sum_a\xi_a^2}\overline{J}_{[n_1,n_2,\ldots,n_m]}(\xi_1^{-1},\xi_2^{-1},\ldots,\xi^{-1}_m)
\ee
These formulas are immediately rewritten via the $\beta$-deformed multiple Borel transforms ${\cal B}_{N,\beta}^{(m)}$ and ${\cal B}_{-\beta N,\beta}^{(m)}$ accordingly. However, these Borel transforms are of much more complicated expressions as compared with (\ref{Borelr}) because of the coefficients ${\cal N}^{-1}_{[n_1,n_2,\ldots,n_m]}$, which make evaluating the infinite sums not that simple.

\section{Conclusion}

Exact solvability is slowly paving its way to the hearts and minds of theorists
working with non-perturbative quantum field theory.
It has its origins in the properties of (functional) integrals \cite{UFN3}:
their independence of the change of integration variables  \cite{MMvir},
and thus is one of the basic features of quantum theory.
The subject is still full of puzzles, including the basic one of how chaotic types of
behavior emerge from exact solvability in the classical limit (see a recent very interesting paper \cite{Son}).
A part of the answer is presumably related to different images which solvability acquires
in different contexts.
There is a very clean part of the story, currently nicknamed {\it superintegrability}
\cite{MMsi}, when all quantum averages are immediately calculable if appropriate
special functions are used, exactly like the motion in quadratic and Newton/Coulomb
potentials, hence the name.
However, so far this is achieved only for Gaussian or hypergeometric (in particular, Selberg) integrals,
and extension requires thorough analysis which will take several steps.

In this paper, we look at the most obvious generalization, a mixture of the
Gaussian and Selberg types, and describe one of the structures which emerges in the answers.
It can look tremendously sophisticated, but this is the usual story: when one
encounters a new type of special functions, one needs time to get used to them and reach
a consensus about appropriate notation and abbreviations.
At this stage, we avoid making simplifications of this type, in order to let people
enjoy and think about the newborn structures in their original form.

Remarkable is a natural appearance of the new variation of Borel transform,
which usually ``improves" infinite series by adding extra factorials \cite{Borel}
or their combinatorial counterparts \cite{MMBorel} in the denominators.
This time the series are not just improved in the asymptotics,
they are {\it cut off} by $\Gamma$-function factors $\frac{N!}{(N-n)!}$
at finite values of $n$ if $B$ is integer, and become polynomials of a finite degree
actually regulated by the size $N$ of the matrix in the underlying matrix model.
Another remarkable fact is that this kind of borelization is applied independently
to all the Miwa variables, thus what we come across is actually
a {\it multiple} {\it enhanced} Borel transform.

Another remarkable fact is that what this Borel acts on are just
the universal exponentials $e^{z_a^2/2}$ multiplied with polynomials,
which do not differ too much from the answers for the Gaussian correlators:
compare (\ref{siG}), (\ref{main}) and (\ref{siG}).
The significance and the degree of generality of this statement,
{\bf the existence of superintegrable direction for deformation of superintegrable theory}
still remains to be appreciated and understood.
What is interesting, however, is that these (\ref{main}) were discovered (see sec.\ref{fromK})
on the base of the theory of strict superintegrability \cite{MMssi,MMskew,CMPT},
which originally could look somewhat artificial, but now proves to be valuable and inspiring.

\section*{Acknowledgements}

Our work is partly supported by the grant of the Foundation for the Advancement of Theoretical Physics ``BASIS" (A. Mir. and A. Mor.) and by the joint grant 21-51-46010-ST-a.


\begin{thebibliography}{12}

\bibitem{MMPS1} A.~Mironov, A.~Morozov, A.~Popolitov and S.~Shakirov,
arXiv:2401.14392

\bibitem{MMMPS3} A.~Mironov et al., to appear

\bibitem{IMM} H.~Itoyama, A.~Mironov, A.~Morozov,
  JHEP {\bf 1706} (2017) 115,
arXiv:1704.08648

\bibitem{MM} A.~Mironov, A.~Morozov,
  Phys.\ Lett.\ {\bf B771} (2017) 503,
arXiv:1705.00976

\bibitem{DiF} P.~Di Francesco, C.~Itzykson, J.~B.~Zuber,
  Commun.\ Math.\ Phys.\  {\bf 151} (1993) 193,
  hep-th/9206090

\bibitem{IdF} P.~Di Francesco, C.~Itzykson,
Ann. Inst. H. Poincare Phys. Theor. \textbf{59} (1993) 117-140,
hep-th/9212108

\bibitem{Ivan1} I.K.~Kostov, M.~Staudacher,
Phys. Lett. \textbf{B394} (1997) 75-81,
hep-th/9611011

\bibitem{Ivan2} I.K.~Kostov, M.~Staudacher, T.~Wynter,
Commun. Math. Phys. \textbf{191} (1998) 283-298,
hep-th/9703189

\bibitem{Orlov} A. Orlov, 
Int. J. Mod. Phys. {\bf A19, supp02} (2004) 276-293, nlin/0209063

\bibitem{MMSh} A.~Mironov, A.~Morozov, S.~Shakirov,
JHEP \textbf{02} (2011) 067,
arXiv:1012.3137

\bibitem{MMShS} A.~Mironov, A.~Morozov, S.~Shakirov, A.~Smirnov,
Nucl. Phys. \textbf{B855} (2012) 128-151,
arXiv:1105.0948

\bibitem{AMMN} A.~Alexandrov, A.~Mironov, A.~Morozov and S.~Natanzon,
JHEP \textbf{11} (2014) 080,
arXiv:1405.1395

\bibitem{Orlov2}  S. Natanzon, A. Orlov, arXiv:1407.8323

\bibitem{Pop} C.~Cordova, B.~Heidenreich, A.~Popolitov, S.~Shakirov,
  Commun.\ Math.\ Phys.\  {\bf 361} (2018)   1235,
  arXiv:1611.03142

  \bibitem{Kaz} V.A. Kazakov, M. Staudacher, T. Wynter, 
hep-th/9601153, 1995 Carg\`ese Proceedings

\bibitem{Ramg} S. Corley, A. Jevicki, S. Ramgoolam, Adv.Theor.Math.Phys. {\bf 5} (2002) 809-839, hep-th/0111222

\bibitem{KPSS} C. Kristjansen, J. Plefka, G. W. Semenoff, M. Staudacher, Nucl.Phys. {\bf B643} (2002) 3-30, hep-th/0205033

\bibitem{BEM} M. Tierz, Mod. Phys. Lett. A19 (2004) 1365-1378, hep-th/0212128\\
A. Brini, B. Eynard, M. Mari\~no, Annales Henri Poincar\'e. Vol. 13. No. 8. SP Birkh\''{a}user Verlag Basel, 2012, arXiv:1105.2012

\bibitem{MKR} R. de Mello Koch, S. Ramgoolam, arXiv:1002.1634

\bibitem{MMten} A.~Mironov, A.~Morozov,
  Phys.\ Lett.\ {\bf B774} (2017) 210,
arXiv:1706.03667

\bibitem{MPS} A.~Morozov, A.~Popolitov and S.~Shakirov,
  Phys.\ Lett. {\bf B784} (2018) 342,
  arXiv:1803.11401

\bibitem{MMsum} A.~Mironov, A.~Morozov,
JHEP {\bf 1808} (2018) 163,
arXiv:1807.02409

\bibitem{MMell}   A.~Mironov, A.~Morozov,
Phys. Lett. \textbf{B816} (2021), 136196,
arXiv:2011.01762; ibid.,
136221,
arXiv:2011.02855

\bibitem{MMkon} A.~Mironov, A.~Morozov,
Eur. Phys. J. \textbf{C81} (2021) 270,
arXiv:2011.12917

\bibitem{MMl} A. Mironov, A. Morozov,  Phys.Lett. {\bf B816} (2021) 136268,  arXiv:2102.01473

\bibitem{Zabz} L.~Cassia, R.~Lodin, M.~Zabzine,
JHEP \textbf{10} (2020) 126,
arXiv:2007.10354

\bibitem{MMPstud} A.~Mironov, A.~Morozov, A.~Popolitov,
Phys. Lett. \textbf{B824} (2022) 136833,
arXiv:2107.13381

\bibitem{WWWZ1} L.~Y.~Wang, R.~Wang, K.~Wu, W.~Z.~Zhao,
Nucl. Phys. \textbf{B973} (2021) 115612,
arXiv:2110.14269

\bibitem{WWWZ2}  R. Wang, F. Liu, C.H. Zhang and W.Z. Zhao,
Eur. Phys. J. {\bf C82} (2022) 902, arXiv: 2206.13038

\bibitem{MMskew} A.~Mironov and A.~Morozov,
J. High Energ. Phys. {\bf 2023} (2023) 116, arXiv:2210.09993

\bibitem{China1}  A.~Mironov, V.~Mishnyakov, A.~Morozov, A.~Popolitov, R.~Wang and W.~Z.~Zhao,
Eur. Phys. J. C \textbf{83} (2023) 377,
arXiv:2301.04107

\bibitem{China2} A.~Mironov, V.~Mishnyakov, A.~Morozov, A.~Popolitov and W.~Z.~Zhao,
Phys. Lett. \textbf{B839} (2023) 137805,
arXiv:2301.11877

\bibitem{China3}  Fan Liu, A. Mironov, V. Mishnyakov, A. Morozov, A. Popolitov, Rui Wang and Wei-Zhong Zhao,
  Nucl.Phys. {\bf B993} (2023) 116283,
  arXiv:2303.00552

\bibitem{MMMZh} A.~Mironov, V.~Mishnyakov, A.~Morozov, A.~Zhabin,
arXiv:2112.11371

\bibitem{MMsi} A.~Mironov, A.~Morozov,
arXiv:2201.12917

\bibitem{MMZ} A.~Mironov, A.~Morozov, Z.~Zakirova,
Phys. Lett.  \textbf{B831} (2022) 137178,
arXiv:2203.03869

\bibitem{MO} V. Mishnyakov, N. Oreshina,  arXiv:2203.15675

\bibitem{MMssi} A.~Mironov and A.~Morozov,
Eur. Phys. J. \textbf{C83} (2023) 71,
arXiv:2206.02045

\bibitem{MMNek} A. Mironov and A. Morozov,
Phys. Rev. \textbf{D106} (2022) 126004,
arXiv:2207.08242

\bibitem{CMPT} C.T.~Chan, V.~Mishnyakov, A.~Popolitov and K.~Tsybikov,
Eur. Phys. J. C \textbf{83} (2023) 1145,
arXiv:2310.02639

\bibitem{Sel} A. Selberg, 
Norsk. Mat. Tidsskr. {\bf 24} (1944) 71-78

\bibitem{Kan} J. Kaneko,
SIAM.J.Math.Anal. {\bf 24} (1993) 1086-1110

\bibitem{Kad2} K.W.J. Kadell,
Adv.Math. {\bf 130} (1997) 33-102\\
K.W.J. Kadell,
Compositio Math. {\bf 87} (1993) 5-43

\bibitem{Mac} I.G. Macdonald, {\it Symmetric functions and Hall polynomials},   Oxford University Press, 1995

\bibitem{Borel} E. Borel, (1899), 
Ann. Sci. \'Ec. Norm. Sup\'er., Series 3, {\bf 16} (1899) 9-131\\
Godfrey Harold Hardy, {\sl Divergent Series}, (1992) [1949], New York: Chelsea\\
Giovanni Sansone, Johan Gerretsen, {\sl Lectures on the theory of functions of a complex variable. I. Holomorphic functions}, (1960), P. Noordhoff, Groningen\\
Ralph P. Boas, Jr. and R. Creighton Buck, {\sl Polynomial Expansions of Analytic Functions (Second Printing Corrected)}, (1964) Academic Press Inc., Publishers New York, Springer-Verlag, Berlin\\
Peter Miller, {\sl Applied asymptotic analysis}, Graduate Studies in Mathematics, {\bf 75}, American Mathematical Society, Providence, RI, 2006\\
Ovidiu Costin, {\sl Asymptotics and Borel summability}, Chapman \& Hall/CRC Monographs and Surveys in Pure and Applied Mathematics, {\bf 141}, CRC Press, Boca Raton, FL, 2009

\bibitem{MMBorel}  A.~Mironov and A.~Morozov,
Phys. Lett. \textbf{B843} (2023) 138037,
arXiv:2302.05903

\bibitem{Mar} M. Mari\~no, 
Fortsch. Phys. {\bf 62} (2014) 455, arXiv:1206.6272\\
I. Aniceto, G. Basar and R. Schiappa, 
Phys. Rept. {\bf 809} (2019) 1, arXiv:1802.10441\\
D. Sauzin, 
Lec. Notes Math. {\bf 2153} (2016), arXiv:1405.0356\\
E. Caliceti, M. Meyer-Hermann, P. Ribeca, A. Surzhykov and U. D. Jentschura, 
Phys. Rept. 446, 1 (2007) arXiv:0707.1596\\
S.~Garoufalidis, J.~Gu and M.~Marino,
Commun. Math. Phys. \textbf{386} (2021) 469-493,
arXiv:2007.10190\\
S.~Garoufalidis and C.~Wheeler,
arXiv:2203.17029\\
M.~Marino, R.~Schiappa and M.~Schwick,
arXiv:2210.13479\\
J.~Gu and M.~Marino,
arXiv:2211.03871

\bibitem{MMNO} A.~Mironov, A.~Morozov, S.~Natanzon and A.~Orlov,
Lett. Math. Phys. \textbf{111} (2021) 124,
arXiv:2012.09847

\bibitem{MMN1} A. Mironov, A. Morozov and S. Natanzon,  Theor.Math.Phys. {\bf 166} (2011) 1-22,
arXiv:0904.4227;  Journal of Geometry and Physics {\bf 62} (2012) 148-155,
arXiv:1012.0433

\bibitem{Fulton} W. Fulton, {\sl Young tableaux: with applications to representation theory and geometry},
LMS, 1997

\bibitem{MMN} A.~Mironov, A.~Morozov and S.~Natanzon,
JHEP \textbf{11} (2011) 097,
arXiv:1108.0885

\bibitem{MMMP1}  A.~Mironov, V.~Mishnyakov, A.~Morozov and A.~Popolitov,
JHEP \textbf{23} (2020) 065,
arXiv:2306.06623

\bibitem{MMS} A.~Mironov, A.~Morozov and S.~Shakirov,
Int. J. Mod. Phys. \textbf{A27} (2012) 1230001,
arXiv:1011.5629

\bibitem{Max} L.~Cassia, R.~Lodin, M.~Zabzine,
JHEP \textbf{10} (2020) 126,
arXiv:2007.10354

\bibitem{MMCal} A.~Mironov and A.~Morozov,
Phys. Lett. \textbf{B842} (2023) 137964,
arXiv:2303.05273

\bibitem{MMMP2} A.~Mironov, V.~Mishnyakov, A.~Morozov and A.~Popolitov,
Phys. Lett. \textbf{B845} (2023) 138122,
arXiv:2307.01048

\bibitem{GMMMO} A. Gerasimov, A. Marshakov, A .Mironov, A. Morozov, A. Orlov,
Nucl.Phys. {\bf B357} (1991) 565

\bibitem{KMMOZ} S. Kharchev, A. Marshakov, A. Mironov, A .Orlov, A. Zabrodin,
Nucl.Phys. {\bf B366} (1991) 569-601

\bibitem{JM} E. Date, M. Jimbo, T. Miwa, J.Phys.Soc.J. {\bf 52} (1983) 388-393

\bibitem{DJKM} E. Date, M. Jimbo, M .Kashiwara and T. Miwa, {\sl Transformation
groups for soliton equations}, RIMS Symp. {\sl "Non-linear integrable
systems -- classical theory and quantum theory"} (World Scientific,
Singapore, 1983)

\bibitem{UT} K. Ueno and K.Takasaki,  Adv.Studies in Pure Math. {\bf 4} (1984) 1

\bibitem{Miwa} T. Miwa, {\sl Proceedings of the Japan Academy}, {\bf 58} (1982)

\bibitem{GKM} S.~Kharchev, A.~Marshakov, A.~Mironov, A.~Morozov, A.~Zabrodin,
  Phys.\ Lett. {\bf B275} (1992) 311,
  hep-th/9111037\\
S.~Kharchev, A.~Marshakov, A.~Mironov, A.~Morozov, A.~Zabrodin,
  Nucl.\ Phys.\ {\bf B380} (1992) 181,
  hep-th/9201013

  \bibitem{Kharchev} S. Kharchev, hep-th/9810091

\bibitem{versus} S. Kharchev, A. Marshakov, A. Mironov, A. Morozov,
Nucl.Phys. {\bf B397} (1993) 339-378, hep-th/9203043

\bibitem{UFN3} A. Morozov,
Phys.Usp.(UFN) {\bf 37} (1994) 1;
hep-th/9502091; hep-th/0502010\\
A. Mironov, Int.J.Mod.Phys. {\bf A9} (1994) 4355; Phys.Part.Nucl.
{\bf 33} (2002) 537; hep-th/9409190

\bibitem{MMvir} F. David, Mod.Phys.Lett. {\bf A5} (1990) 1019\\
A. Mironov, A. Morozov, Phys.Lett. {\bf B252} (1990) 47-52\\
J. Ambj{\o}rn, Yu. Makeenko, Mod.Phys.Lett. {\bf A5} (1990) 1753\\
H. Itoyama, Y. Matsuo, Phys.Lett. {\bf 255B} (1991) 20

\bibitem{Son} S.~Negro, F.K.~Popov and J.~Sonnenschein,
Phys. Rev. \textbf{D108} (2023) 105024,
arXiv:2211.14150

\end{thebibliography}
\end{document}